# Anderson Localization, Non-linearity and Stable Genetic Diversity

Charles L. Epstein*
Department of Mathematics
and
Laboratory for Structural NMR Imaging
University of Pennsylvania

February 28, 2006

**Abstract**

In many models of genotypic evolution, the vector of genotype populations satisfies a system of linear ordinary differential equations. This system of equations models a competition between differential replication rates (fitness) and mutation. Mutation operates as a generalized diffusion process on genotype space. In the large time asymptotics, the replication term tends to produce a single dominant quasispecies, unless the mutation rate is too high, in which case the populations of different genotypes becomes de-localized. We introduce a more macroscopic picture of genotypic evolution wherein a random replication term in the linear model displays features analogous to Anderson localization. When coupled with non-linearities that limit the population of any given genotype, we obtain a model whose large time asymptotics display stable genotypic diversity.

# Introduction

This paper contains a proposal for a class of theories of genotypic evolution that display stable, arbitrarily complex genetic diversity. Our models are built out of pieces that have been on the shelves for quite a while, but perhaps have not before been placed together.

*Research partially supported by DARPA under the FUNBIO program.

We start out with linear "spin glass" models, which cast genotypic evolution as competing processes of replication and mutation. We posit the existence of sequences, $\{R_n\}$, of replication (or fitness) matrices so that the combined replication-mutation system exhibits properties, in the thermodynamic limit of large genome length, analogous to Anderson localization. Such a model already exhibits a weak form of genetic diversity, having a large number of well defined, well separated, "long-lived" quasi-species. In a linear model it is more or less inevitable that, in the long run, either a single species comes to dominate, or localization breaks down and there are no well defined quasispecies. Our additional step is to add a quadratic term that limits the growth of any given genotype. This step is suggested by a paper of Nelson and Shnerb, where they show, in a continuum population biology model, that such a term, when coupled to Anderson localization does in fact lead to an asymptotic state with stable diversity, see [8].

The main contribution of this paper is to put these two pieces together in the context of genotypic evolution and suggest potentially fruitful directions for further research in both evolution and spectral theory. In small numerical examples we show that our localization hypothesis is not unreasonable. The rigorously established results showing that localization occurs for Schrödinger operators with sufficiently weak diffusion, and that, when it occurs, is generic, gives support for the idea that such models should exist and that the localization property should be insensitive to the details of the model. Finally, there is a certain sublime beauty to a world in which the *randomness* of the mapping from genotype to fitness conspires with environmental *limitations* on population size to produce stable genetic diversity.

Acknowledgments

I would like to thank Ben Mann for insisting that I participate in the FUNBIO program and securing the funding to make it happen. I am very thankful to all participants in the DARPA FUNBIO workshops and most especially to Richard Lenski, Sally Otto, Michael Deem, Jeong-Man Park, Chris Adami, and Jack Morava for sharing their ideas on biology and evolution with me. I would also like to thank Michael Deem for several useful suggestions for improvement of an earlier drafts. Finally I am very grateful to Harvey Rubin for telling me the biochemical facts of life and to John Schotland for his considerable help with statistical mechanics and many useful suggestions related to this work.

# 1 Linear models

Recently there has been a great deal of interest in the connections between various models that arise in statistical mechanics and models of genetic evolution. Early models were defined by Eigen and Crow-Kimura, see [5, 4]. For a good survey of this subject with many references to the literature see [7]. Genotypes are described as sequences $(s_1, \ldots, s_n)$, where the entries $\{s_j\}$ are drawn from a finite alphabet. For example, if one wishes to model chromoso-mal evolution the alphabet is that of nucleotides $\{A, C, G, T\}$ (or, for RNA, $\{A, C, G, U\}$). If one wishes to studies protein evolution, then one might use the list of the 20 amino acids. In the interest of simplicity, most investigators simply use a two letter alphabet, which can be thought of as purines and pyrimidines. We let $\mathcal{G}_{n,l}$ denote the set of possible genotypes of length $n$ expressed in the given fixed alphabet with $l$ members. If all genotypes are possible, then $|\mathcal{G}_{n,l}| = l^n$. In the sequel we let $N = l^n$. The different genotypes can therefore be labeled by the set of integers $\mathcal{I}_{n,l} = \{1, 2, \ldots, N\}$, though this labeling scheme conveys no further information.

We specify a model for mutation from one genotype to another by assigning probabilities $\{m_{ij} : i \neq j \in \mathcal{I}_{n,l}\}$ that, in a given unit of time, the genotype $S_i$ mutates to the genotype $S_j$. If we think of $\mathcal{G}_{n,l}$ as the vertices of a directed graph, then we add a directed edge from $S_i$ to $S_j$ if $m_{ij} > 0$. Let $\boldsymbol{P}(t) = (P_1(t), \ldots, P_N(t))$, where $P_j(t)$ is the population of the genotype $S_j$ at time $t$. In addition to mutation, each genotype has a replication rate, $r_i$ so that, in the absence of mutation we would have the simple differential equation describing the change of the population of genotype $S_j$. :

$$\frac{dP_i}{dt} = r_i P_i(t). \tag{1}$$

Hence $r_i$ is the difference of the birth and death rates for the genotype $S_i$. The *replication* matrix, $R$, is defined to be

$$R_{ij} = \begin{cases} 0 \text{ if } i \neq j \\ r_i \text{ if } i = j. \end{cases} \tag{2}$$

This matrix is often called the "fitness" matrix, but following a suggestion of Michael Deem, we use the more precise term "replication" matrix.

The mutational process is described by the *mutation* matrix, $M$, given by:

$$M_{ij} = \begin{cases} m_{ji} & \text{if } i \neq j \\ -\sum_{k \neq i} m_{ik} & \text{if } i = j. \end{cases} \tag{3}$$

The negative diagonal term is required so that the total mutational flux out of a given genotype, is balanced by an equal decrease in its population. The standard linear model for the time course of the genotype *populations* is then

$$\frac{d\boldsymbol{P}}{dt} = R\boldsymbol{P} + M\boldsymbol{P}. \tag{4}$$

In many prior papers on this subject, the model is described as a model for population *densities*, rather than the populations themselves. These models are equivalent, under a simple change of variables, to a linear model, and it is the linear model that is amenable to analysis. For many choices of $R$ and $M$, these models have convenient representations in terms of Pauli spin matrices, which in turn allows the application of techniques developed to study spin glass models in statistical mechanics.

In most of the previous analyses of these models, the alphabet has 2 letters. If $S_i = (s_1, \dots, s_n)$, $S_j = (s'_1, \dots, s'_n)$ are two genotypes, then the Hamming distance between them equals the number of entries where they differ. If we represent the genotypes as sequences of plus and minus ones, then

$$d_H(S_i, S_j) = \frac{1}{2}\left[ n - \sum_{k=1}^{n} s_k s'_k \right]. \tag{5}$$

If our alphabet has $l$-letters, then more generally we can define a metric by first defining a metric on the alphabet: let $\mathscr{A} = \{a_1, \dots, a_l\}$ then $d_{\mathscr{A}} : \mathscr{A} \times \mathscr{A} \to [0, \infty)$ is a function that satisfies:

$$\begin{aligned} &d_{\mathscr{A}}(a_i, a_j) \geq 0 \text{ and equals 0 only if } i = j. \\ &d_{\mathscr{A}}(a_i, a_j) = d_{\mathscr{A}}(a_j, a_i) \text{ for all pairs } i, j. \\ &d_{\mathscr{A}}(a_i, a_j) \leq d_{\mathscr{A}}(a_i, a_k) + d_{\mathscr{A}}(a_k, a_j) \text{ for all triples } i, j, k. \end{aligned} \tag{6}$$

The Hamming metric on $\mathscr{G}_{n,l}$ is then defined by

$$d_{H,\mathscr{A}}(S_i, S_j) = \sum_{k=1}^{n} d_{\mathscr{A}}(s_k, s'_k). \tag{7}$$

In earlier papers, which consider an alphabet with two letters, the mutation probabilities are often taken to be functions of the Hamming distance. In the paramuse model the mutation matrix is specified by

$$M_{ij} = \begin{cases} \mu & \text{if } d_H(i, j) = 1 \\ -n\mu & \text{if } i = j \\ 0 & \text{otherwise.} \end{cases} \tag{8}$$

The probability-per-unit-time of changing one letter is $\mu$ and the probability of changing more than one letter is zero. In the Eigen model, $m_{ij}$ is a function of the Hamming distance between $i, j$ of the form

$$m_{ij} = \mu^{d_H(S_i,S_j)}(1-\mu)^{n-d_H(S_i,S_j)}. \tag{9}$$

The assumption here is that probability-per-unit-time of mutation at each site on the genome is equal to that of every other, and that they are also independent of one another. In this case the division between mutation and replication is not as simple. In our subsequent analysis we stick to the representation given in (4). The precise nature of $M$ is less important than the assumption that the semi-group $e^{tM}$ should have the qualitative properties of a diffusion process:

1. It should be positivity improving, i.e. if $\boldsymbol{P}_0$ is a vector with non-negative coefficients, then $e^{tM}\boldsymbol{P}_0$ has positive coefficients.

2. It should decay very rapidly as we depart from the diagonal.

These conditions amount to the requirements that the off-diagonal entries of $M$ are non-negative and rapidly decaying as we depart from the diagonal.

The population vector $\boldsymbol{P}$ defines a function on the vertices of the genotype graph $\mathcal{G}_{n,l}$. In these models, a quasi-species is represented by a population vector that is highly concentrated around a single vertex or small cluster of vertices. A population vector with several such clusters, which are well separated on the graph, would represent a collection of quasi-species. The mutational process is a diffusion on this graph. The eigenvectors of the matrix $M$ are not localized. The Perron-Frobenius Theorem implies that the largest eigenvalue of $M$, $\lambda_P$, corresponds to an eigenvector $v_P$ all of whose entries are positive. Indeed, usually $v_P = (1, \ldots, 1)$. Under the effects of mutation *alone*, any initially non-negative population, $\boldsymbol{P}_0$, behaves asymptotically as

$$\boldsymbol{P}(t) \sim \langle \boldsymbol{P}_0, v_P \rangle v_P e^{\lambda_P t}. \tag{10}$$

So the effect of the mutational process is to smear out the population and destroy any localized populations that might be present in the initial distribution of genotypes.

On the other hand, the replication matrix $R$ is diagonal and if it has distinct entries, then each basis vector $e_j = (0, \ldots, 0, 1, 0, \ldots, 0)$ (1 in the $j$th place) defines a quasi-species. In the absence of mutation, the population of this quasi-species evolves as $e^{r_j t} P_j(0)$. Indeed the matrices $\{e^{t(R+M)} : t > 0\}$ are also positivity improving and hence have a positive Perron-Frobenius eigenvector, $v_P$, corresponding to the largest eigenvalue, $\lambda_P$. From our perspective, the problem here is that, in the long time limit, the population will again satisfy (10) and be dominated by the population distribution (quasi-species or not) with this highest replication rate. For most of the analyses of these models this was not really viewed a difficulty, as the problem under analysis was the stability of a single quasi-species under various levels of mutation, and for various choices of smooth replication landscape.

By smooth we mean that the replication rate is smooth as function of the genotype with respect to the distance function defined on $\mathcal{G}_{n,l}$. This sort of analysis can be regarded as focusing on a small neighborhood of a vertex in $\mathcal{G}_{n,l}$. If we assume that $m << n$ sites participate in the evolutionary process, then the analysis proceeds on $\mathcal{G}_{m,l}$ viewed as a subgraph of $\mathcal{G}_{n,l}$. On this subgraph (length scale), even a macroscopically random replication landscape could well appear quite smooth. Hence, thermodynamic analyses, like that in [11], can be viewed as genotypically localized, short time analyses that take place within the larger macroscopic framework of genetic evolution.

In this paper we consider questions related to the long time macroscopic structure of genotypic evolution. We focus on aspects of linear models that are connected to the randomness of the replication rate matrix, and beyond that on the consequences of non-linear corrections that are needed to account for the finiteness of resources. Our ideas related to randomness and localiza-tion are inspired by the seminal work of P.W. Anderson [2], and the effects of non-linear corrections, by the work of D.R. Nelson and N.M. Shnerb in population biology, see [8].

## 2 Anderson Localization

The combined linear model given in (4) represents a competition between the replication term, which, if the diagonal entries are random, tends to preserve quasi-species, and the mutation term, which tends to destroy them. As such, these models have a great deal in common with the models for conduction in semiconductors studied by Anderson. In his seminal work and many subsequent analyses, it has become clear that there is a very fundamental and generic localization property shared by systems with a random "potential." Before proceeding with out discussion, we briefly discuss the analysis of continuum models of the form:

$$\begin{aligned} u_t(x,t) &= (L+E)u(x,t) \text{ where} \\ Lu(x,t) &= \mu\Delta u(x,t) + q(x)u(x,t), \end{aligned} \tag{11}$$

with $x \in \mathbb{R}^p$, $t \in [0,\infty)$, and $E$ a positive constant. It is well known that these equations are positivity improving: if $u(x,0) \geq 0$ for all $x$, then $u(x,t) > 0$ for all $x$ and $t > 0$, see [10]. We are therefore free to interpret $u(x,t)$ as the density of the population located at position $x$ at time $t$.

While it is very difficult to see how an evolutionary model, with underlying space a graph like $\mathcal{G}_{n,l}$, can be approximated by a continuum model like that in equation (11), there are strong structural analogies between this type of evolution equation and (4). Under time evolution, the first term, $\mu\Delta u$, generates a spatially homogeneous diffusion process. As time goes to infinity, this term will lead to a spatially uniform population. The multiplication operator $q(x)$ is analogous to the replication matrix. We include $E$ to have a background environmental energy or temperature in the problem.

Formally, the solution to (11) is given by

$$u(x,t) = e^{tE}e^{tL}u(x,0). \tag{12}$$

The qualitative behavior of the solution is determined by the spectral theory of $L$. There are three simple possibilities (and many cases where the answer is not known). If $q$ decays, sufficiently rapidly, as $\|x\|$ tends to infinity, then usually the spectrum of $L$ consists of purely absolutely continuous spectrum in $(-\infty, 0]$, along with some $L^2$-eigenstates with positive eigenvalues. There are, at most, finitely many $L^2$-eigenstates with eigenvalues in an interval of the form $[\epsilon, M]$, where $\epsilon > 0$, though the positive spectrum can accumulate at 0. If there are no $L^2$-eigenstates then a localized initial condition spreads out as $t \to \infty$. If there are $L^2$–eigenstates, then the corresponding eigenvectors usually have large overlaps in their supports. If there is a maximum positive eigenvalue $\lambda_0$, with a positive eigenvector, $\psi_0$, (a vacuum state), then asymptotically the solution behaves like $\langle\psi_0, u(\cdot, 0)\rangle\psi_0 e^{(E+\lambda_0)t}$.

Another simple possibility is that $q(x)$ tends to infinity as $\|x\|$ tends to infinity. In this case the spectrum is pure point spectrum,

$$\{\lambda_0 > \lambda_1 \geq \lambda_2 \geq \ldots\},$$

with each eigenvalue of finite multiplicity, and $\lim_{j\to\infty}\lambda_j = -\infty$. Let $\{\psi_j\}$ be the eigenstates. These are localized functions, but typically their supports have considerable overlap. An important special case is given by the $q(x) = \|x\|^2$, the harmonic oscillator. The eigenfunctions are of the form $p_j(x)e^{-\frac{1}{2}\|x\|^2}$, where $p_j(x)$ are polynomials. In the large time limit, the solution again behaves like $\langle\psi_0, u(\cdot, 0)\rangle\psi_0 e^{(E+\lambda_0)t}$. Since $q$ is unbounded, this case would not appear to have much to do with the evolutionary models above. These two cases are extensively described in [10].

The third case is that $q$ remains bounded but has no asymptotic or periodic behavior as $\|x\| \to \infty$. These are what are often referred to as "random potentials" in the mathematics literature. A simple example would be an almost periodic function like $\cos x + \cos\sqrt{2}x$. In this case the spectrum of $L$ can behave in a very remarkable way. In his seminal 1958 paper, [2, 13], Anderson argued that, when $\mu$ is not too large, the operator $L$ has dense point spectrum lying in intervals, and the corresponding eigenfunction are exponentially localized. Though it required the development of considerable analytic technique, the substance of these assertions has been verified in many special cases. In one and two dimensions (i.e. $x \in \mathbb{R}$ or $\mathbb{R}^2$) it has been shown that, with probability one (with respect to the choice of potential), the spectrum of $L$ is dense pure point spectrum on a half line $\{\lambda_j\} \subset (-\infty, \lambda_0]$ and the corresponding eigenfunctions $\{\psi_j\}$ fall off exponentially. That is, the closure of the set $\{\lambda_j\}$ is the half line $(-\infty, \Lambda_0]$, and

$$\begin{aligned} L\psi_j &= \lambda_j\psi_j \\ |\psi_j(x)| &\leq C_j e^{\frac{-|x-x_j|}{\xi_j}}. \end{aligned} \tag{13}$$

In higher dimensions, more complicated things can happen. For example, one could have dense point spectrum in an interval $[\Lambda_1, \Lambda_0]$, and then an interval of continuous spectrum $[\Lambda_3, \Lambda_2]$. Nonetheless the appearance of intervals of dense point spectrum is a generic property for many classes of potentials. Discrete models, analogous to (11), defined for functions on the lattices have also been extensively studied. For sufficiently weak diffusion, Anderson localization has also been shown to be a generic property. See [6, 9, 3] for mathematical results in this field and further references.

The third case seems to be closest to what is expected of a realistic replication landscape: the replication rate is constantly varying throughout genotype space and is neither periodic nor has any asymptotic behavior. This does not preclude the replication landscape from being locally smooth or having large regions where it is approximately constant. We let $\{L_n = R_n + M_n\}$ denote a sequence of operators acting on functions on $\mathcal{G}_{n,l}$. In our subsequent analysis we use the following **localization ansatz:**

> As $n$ tends to infinity, the spectrum of the operators $L_n$ becomes dense in some interval with right end point $\sup \operatorname{spec}(L_n)$. The corresponding normalized eigenvectors become exponentially localized, with the overlaps in support uncorrelated to the differences in energy.

One might want to suppose that the matrices $\{R_n\}$ converge to a infinite diagonal matrix $R_\infty$ and the sequence of discrete diffusion operators $\{M_n\}$ converge, in some sense, to an operator acting on $\ell^2$. It is by no means obvious how to normalize the sequence $\{M_n\}$ so that the limit produces a non-trivial diffusion process. At realistic mutation rates, genotype space seems to be explored very slowly, so there is not much practical difference between a very large, but finite length genome, and an infinite length genome.

In the random case, the degree of overlap of eigenvectors should not be correlated to the difference in energies, i.e. if $\lambda_i$ and $\lambda_j$ are nearby eigenvalues it is highly unlikely that the supports of the corresponding eigenvectors $\psi_i$ and $\psi_j$ have a substantial overlap. If the diagonal matrix $R_n$ has distinct and say strictly monotonically increasing entries, then it is again the case that, for small enough $\mu$, the eigenvectors of $L_n$ are highly localized. What distinguishes this case from the random case is that now the overlap in the eigenvectors is highly correlated with the difference in energy: if $|\lambda_i - \lambda_j|$ is small then is very likely that the supports of $\psi_i$ and $\psi_j$ have a large overlap. This becomes quite important when we consider the effects of the non-linear corrections.

To the best of my knowledge, this precise situation has not yet been analyzed, though considerable effort has been devoted to studying analogous questions on the lattices $\mathbb{Z}^d$, and Anderson localization has been rigorously established in many representative cases, see [9]. In the physics literature it has been shown that Anderson Localization occurs for a system based on the Bethe lattice, see [1]. This is of interest for us, as the Bethe lattice embeds isometrically into the hyperbolic plane. Hence this indicates that the more efficient diffusion that occurs in negatively curved spaces does not destroy Anderson Localization. In the final section we give some numerical examples suggesting that this sort of localization does occur on the graphs $\mathcal{G}_{n,l}$.

Anderson localization is a phenomenon that exhibits phase transitions: for sufficiently small diffusion or at sufficiently low energies it can be expected to occur, but as the diffusion rate or energy become too large it may abruptly disappear, with the pure point spectrum being replaced by continuous spectrum. Such a transition would have interesting consequences for the underlying genetic system.

*Remark* 1. There is a somewhat different infinite $n$ limit of genotype space that may be more appropriate than simply taking the genome length to infinity. We could also consider the space

$$\mathcal{G}^s_{\infty,l} = \bigsqcup_{n=1}^{\infty} \mathcal{G}_{n,l}. \tag{14}$$

The space $\mathcal{G}^s_{\infty,l}$ contains genotypes of all different lengths, and provides a framework for studying interactions, i.e. recombination and splicing, among the genetic material of very different types of organisms, e.g. viruses and eukaryotes. It would seem an important question to understand under what mutational structures, generic random replication landscapes exhibit localization on $\mathcal{G}^s_{\infty,l}$. It may also provide a framework where it is easier to take the thermodynamic limit of the mutation process.

# 3 Weak genetic diversity

Before considering the role of non-linearities, we consider what a linear model satisfying the localization ansatz would predict. Let us fix a large value of $n$ so that $L_n$ has many exponentially localized, well separated eigenstates near the supremum of the spectrum. Indeed we normalize so that

$$\sup \operatorname{spec}(L_n) = 0. \tag{15}$$

We follow the continuum model and explicitly include an energy, which could represent an environmental temperature, in our system:

$$\frac{d\boldsymbol{P}}{dt} = (L_n + E)\boldsymbol{P}. \tag{16}$$

In a linear model the addition of $E$ has no qualititative effect on the solution, it simply scales the result by $e^{tE}$. As we shall see, this is no longer the case once we include non-linear corrections.

The localization ansatz is that, near to 0, there is a large number of eigenvalues $\{0 = \lambda_0 \geq \lambda_1 \geq \dots\}$, such that the corresponding normalized eigenvectors $\{\psi_\alpha\}$ are highly localized, and the overlaps in their supports are uncorrelated with their energy differences. Quantitatively we take this to mean that for each $\alpha$, there is a $j_\alpha \in \mathcal{G}_{n,l}$ and positive numbers $\xi_\alpha$, $C_\alpha$, so that

$$\psi_\alpha(j) < C_\alpha e^{-\frac{d_H(j.j_\alpha)}{\xi_\alpha}}, \tag{17}$$

and for $\alpha \neq \beta$

$$\sum_{j\in\mathcal{G}_{n,l}} |\psi_\alpha(j)\psi_\beta(j)| << 1, \tag{18}$$

with high probability, especially if $|\lambda_\alpha - \lambda_\beta|$ is small. Moreover, we assume that $\{\psi_\alpha(j)\}$ is positive for most values of $j$. Because of exponential localization, this assumption does not contradict the fact that the eigenvectors are an orthonormal set.

If $\boldsymbol{P}_0$ is an initial population distribution, then evolving under equation (16), the population satisfies:

$$\boldsymbol{P}(t) = \sum_\alpha \langle \boldsymbol{P}_0, \psi_\alpha \rangle \psi_\alpha e^{(E+\lambda_\alpha)t}. \tag{19}$$

For long times only the terms with $E + \lambda_\alpha > 0$ make a significant contribution to $\boldsymbol{P}(t)$. The $\lambda_0$-term is still the dominant term, but there may be many terms with $\lambda_0 - \lambda_\alpha$ quite small, which therefore make significant contributions for a long time. Because the eigenstates are well localized and well separated, there can well be different dominant terms at different locations in genotype space. Thus, even without non-linear corrections, a model satisfying the localization ansatz would exhibit some sort of genetic diversity, which we call *weak genetic diversity*.

# 4 Non-linear effects

Nelson and Shnerb modify the model in (11) by adding a non-linear term:

$$u_t(x,t) = (L+E)u(x,t) - bu^2(x,t), \tag{20}$$

where $b > 0$. The effect of this term is to limit $u(x,t)$ to remain less than $E/b$. More generally if $b$ is replaced by any positive $L+E$-super-harmonic function, $B(x)$, so that, for all $x$,

$$(L+E)B(x) - B^3(x) < 0, \tag{21}$$

then the maximum principle shows that if initial data $u(x,0) < B(x)$ for all $x$, then $u(x,t) < B(x)$ for all $x$ and $t > 0$.

The remarkable observation made by Nelson and Shnerb is that, if $L$ exhibits Anderson localization, then the large time asymptotics of the non-linear equation actually depend on all the eigenvalues of $L$ with $\lambda_\alpha + E > 0$. If we let

$$c_\alpha(t) = \langle u(\cdot,t), \psi_\alpha \rangle, \tag{22}$$

then, under (20), they evolve according to:

$$\frac{c_\alpha(t)}{dt} \simeq (E+\lambda_\alpha)c_\alpha(t) - w_\alpha c_\alpha^2(t), \tag{23}$$

where

$$w_\alpha = b \int \psi_\alpha^3(x)dx. \tag{24}$$

Because the eigenstates are highly localized, the coefficients for the cross terms

$$b \int \psi_\alpha(x)\psi_\beta(x)\psi_\gamma(x)dx \tag{25}$$

are very small, and that is why they can be ignored. We can solve (23) to obtain:

$$c_\alpha(t) = \frac{c_\alpha(0)e^{(E+\lambda_\alpha)t}}{1 + c_\alpha(0)\frac{w_\alpha}{E+\lambda_\alpha}(e^{(E+\lambda_\alpha)t} - 1)}. \tag{26}$$

Hence as $t \to \infty$, all species such that $E + \lambda_\alpha > 0$ and $c_\alpha(0) > 0$ have a finite, non-zero asymptotic value given by

$$\lim_{t\to\infty} c_\alpha(t) = \frac{E + \lambda_\alpha}{w_\alpha}. \tag{27}$$

Here we see the importance of the "$E$-term" in a non-linear model.

We can apply similar considerations to a somewhat larger class of models, in which we include a second non-linearity to limit the total population

$$u_t(x,t) = (L + E)u(x,t) - B(x)u^2(x,t) - pu(x,t)\int u(x,t)dx. \tag{28}$$

The second, non-local term, can be shown to impose a limit on the total population $\int u(x,t)dx$, which the first term does not do. One can also show that, so long as $B(x) > c > 0$, the asymptotic behavior of a non-negative solution of (28) is similar to that for (20). If $B \equiv 0$, then the model in (28) has many critical points, none stable and none with a large number of non-zero coefficients.

The analogue of the model in (20) is an equation of the form:

$$\frac{d\boldsymbol{P}(t)}{dt} = (L + E)\boldsymbol{P}(t) - b\boldsymbol{P}(t). * \boldsymbol{P}(t). \tag{29}$$

Here we use the MATLAB notation for component-wise vector multiplication: if $\boldsymbol{v} = (v_1, \dots, v_m)$ and $\boldsymbol{w} = (w_1, \dots, w_m)$, then

$$\boldsymbol{v}. * \boldsymbol{w} = (v_1w_1, v_2w_2, \dots, v_mw_m). \tag{30}$$

For this discussion we take $M$ given by (3), though much of what we say should remain true with any reasonable choice of $M$. Let $\boldsymbol{B}$ be a positive super-harmonic vector $(L + E)\boldsymbol{B} - \boldsymbol{B}. * \boldsymbol{B}. * \boldsymbol{B} < 0$. If we replace (29) with

$$\frac{d\boldsymbol{P}(t)}{dt} = (L + E)\boldsymbol{P}(t) - \boldsymbol{B}. * \boldsymbol{P}(t). * \boldsymbol{P}(t), \tag{31}$$

and $0 \le P_j(0) < B_j$ for all $j$, then $0 \le P_j(t) < B_j$ for all $j$ and $t > 0$. A model similar to (28) is given by

$$\frac{d\boldsymbol{P}(t)}{dt} = (L + E)\boldsymbol{P}(t) - \boldsymbol{B}. * \boldsymbol{P}(t). * \boldsymbol{P}(t) - p\langle \boldsymbol{P}, \boldsymbol{1}\rangle \boldsymbol{P}, \tag{32}$$

where $p > 0$ and $\mathbf{1} = (1, \ldots, 1)$. These models are all positivity preserving. Their basic mathematical properties are established in the Appendix. For simplicity we proceed with the model given in (29).

As noted, the solution operator for equation (29) is positivity preserving. If

$$c = \frac{\sup\{E + r_j\}}{b}, \tag{33}$$

then a simple maximum principle argument shows that, if $0 \leq P_j(0) < c$, for all $j$, then $0 < P_j(t) < c$ for all $t > 0$. If $L$ satisfies the localization ansatz, then the analysis used to derive (23)–(27) applies, *mutatis mutandis* to (29). The spectrum, $\{\lambda_\alpha\}$, of $L$ is quite dense near to zero, and the corresponding eigenvectors, $\{\psi_\alpha\}$ are highly localized. As before we express the initial data as

$$\boldsymbol{P}(0) = \sum_\alpha \langle \boldsymbol{P}(0), \psi_\alpha \rangle \psi_\alpha = \sum_\alpha c_\alpha(0)\psi_\alpha. \tag{34}$$

We need to express the non-linear term in the eigenbasis:

$$w_{\alpha\beta,\gamma} = \langle \boldsymbol{B}. * \boldsymbol{P}(t). * \boldsymbol{P}(t), \psi_\gamma \rangle = b \sum_j \sum_{\alpha,\beta} c_\alpha(t) c_\beta(t) \psi_\alpha(j) \psi_\beta(j) \psi_\gamma(j). \tag{35}$$

The localization ansatz implies that in general, unless $\alpha = \beta = \gamma$, $w_{\alpha\beta,\gamma}$ is very small, especially if $|\lambda_\alpha - \lambda_\beta|$ is small. We set $w_\alpha = w_{\alpha\alpha,\alpha}$. Hence, the coefficients again satisfy (23) and therefore the solution is again given by (26), with long time asymptotics given by (27),

$$\lim_{t \to \infty} c_\alpha(t) = \frac{E + \lambda_\alpha}{w_\alpha}. \tag{36}$$

Thus we see that coupling localization in genotype space with a simple non-linearity produces a model exhibiting long time genetic diversity. We get a large collection of distinct quasi-species occupying different parts of genotype space.

What distinguishes a random replication matrix from one with monotonely increasing diagonal entries is that, in the latter case, the eigenvectors with large overlaps in their supports tend to have nearby energies. This means that, when the non-linear terms are included, paths exist in the energy landscape defined by the spectral theory of $L$ that give the population the opportunity to cascade toward states with lower energy. This claim is born out by the numerical simulations of the non-linear equation in the next section. In the random case no such paths exist and this further supports our claim that these models will display stable genetic diversity. It also suggests a connection between these models and an interesting percolation problem on an energy landscape defined by the spectral theory of $L$.

There are a variety of other interesting phenomena that could be obtained with models satisfying the localization ansatz. For example, if $L$ has a band, lying below the localized states, of eigenvalues with *non*-localized eigenstates, then an increase in $E$ to $E'$, would require re-expressing the data in terms of the new eigenstates with $\lambda_\alpha + E' > 0$, which would include some states with non-localized eigenvectors. Evolution at this higher energy could result in significant shuffling of the genotype populations. If the energy subsequently dropped back to a range where the relevant eigenstates are again localized, then the system would eventually settle into a steady state with considerable genetic diversity, possibly quite different from the state we had prior to the temporary increase in energy.

# 5 Numerical examples

In this section we present some numerical evidence for the localization ansatz. In a variety of papers, notably in [11], it is shown that, in order for quasispecies to exist in the large $n$ limit, it is necessary that $n\mu$ be less than the maximal diagonal term in the replication matrix. For our numerical simulations we divide by $n$ so that $\mu$ is fixed and

$$M_{ij} = \begin{cases} -\mu & \text{if } i = j \\ \frac{\mu}{n} & \text{if } d_H(i, j) = 1 \\ 0 & \text{otherwise.} \end{cases} \tag{37}$$

We then compute the eigenvalues and eigenvectors of matrices of the form $L = R + M$, where $R$ is a diagonal matrix with (pseudo)random, uniformly distributed entries, scaled to lie in $[0, 1]$. Physically this amounts to replacing the time parameter $t$ by $t/n$. For purposes of comparison, we also consider diagonal matrices with distinct but smoothly growing entries, e.g. $R_{ii} = \tanh(2^{-n} i)$, and replication matrices arising in single peak fitness landscapes, $R_{ii} = 1/(1 + d_H(S_i, S_0))$. Our numerical experiments display several striking phenomena.

For a given matrix $L$, let $\{\lambda_\alpha\}$ denote the spectrum and $\{\psi_\alpha(k)\}$, the corresponding normalized eigenvectors. The eigenvalues are indexed in increasing order. In our experiments we see that, with a random replication matrix, the spectrum is distributed fairly uniformly over an interval, with decreasing density near the upper endpoint. This is in agreement with known results on the spectral density function in the continuum case, see [2, 13]. The matrices we consider are symmetric, so the eigenvectors are real and orthonormal:

$$\sum_{k=1}^{2^n} \psi_\alpha(k)\psi_\beta(k) = \delta_{\alpha\beta}. \tag{38}$$

To measure the extent of the overlap in the support of the eigenvectors we compute the following sums

$$C_{\alpha\beta} = \sum_{k=1}^{2^n} |\psi_\alpha(k)\psi_\beta(k)|. \tag{39}$$

If the eigenvectors were perfectly localized then the matrix, $C_{\alpha\beta} - \delta_{\alpha\beta}$, would be zero. If the eigenvectors were completely de-localized, like $(2^{-\frac{n}{2}} e^{\frac{2\pi ijk}{2^n}} : k = 0, \ldots, 2^n - 1)$, then $C_{\alpha\beta} = 1$ for all $\alpha, \beta$. A more compact measure of mean localization is provided by the averages along rows:

$$C_\alpha = \frac{1}{2^n} \sum_{\beta \neq \alpha} C_{\alpha\beta}. \tag{40}$$

Figures 1–2 show the results of simulations for $n = 7$ and 11. The upper row of each figure is a surface plot of $\log_{10} C_{\alpha\beta}$ (suitably cutoff from below) for a random replication matrix (on the left) and for $R_{ii} = \tanh(2^{-n} i)$ (on the right). As noted above, the random model has a uniformly distributed diagonal scaled to lie in the interval [0, 1]. We use the tanh-function, which has well defined asymptotics, to avoid harmonic oscillator-like behavior. For these examples $\mu = 10^{-5}$.

The lower rows of these figures are plots of $\log_{10}(2^{-n} \sum_\beta C_{\alpha\beta})$. We see that, as $n$ grows, the models with random replication matrices tend strongly toward localization. Moreover, the overlaps tend to be quite random and uncorrelated with energy differences. If $R_{ii} = \tanh(2^{-n} i)$, then the eigenvectors also tend to be rather localized, though not as strongly as in the random case. However the pattern of overlap is highly correlated with the energy difference, and entirely different from the random case. In Figure 3 we show the same data, with $n = 11$ and $\mu = 10^{-5}$, but this time using a single peak replication matrix: $R_{ii} = 1/(1 + d_H(S_i, S_0))$, with $S_0 = (1, \ldots, 1)$, as the "smooth" model. In this case the spectrum of the smooth model is highly degenerate. We again show both plots as the random model is computed with a different realization of $R$. In this case the eigenvectors of the smooth model are highly correlated.

In Figure 4 we show the spectra of $L$ for the examples considered in Figures 2 and 3. For the random model, the spectrum is dense in an interval, and thins out toward the endpoints. The monotone model also has a dense spectrum, whereas the one-peak model has a highly degenerate spectrum. Figure 5 shows the first ten eigenvectors of a random model with $n = 11$ and $\mu = 10^{-5}$. This largely bears out our claim that the overlap in the eigenvectors is uncorrelated with the energy difference, though, in this example, eigenvectors 2 and 4 have considerable overlap. In a similar plot (not shown) using the monotone model, all 10 eigenvectors are located at the extreme left edge and cannot be visually distinguished. Figure 6, shows the effect on the mean overlap in the eigenvalues as $\mu$ is decreased, while $n$ is kept fixed. We see that the mean overlap is roughly proportional to $\mu$. In Figures 1–2(c) we see that the mean overlaps for $n = 7, 9, 11$ are $3 \times 10^{-6}$, $1 \times 10^{-6}$, $3 \times 10^{-7}$ respectively, indicating that as the genome length increases the degree of localization of the eigenvectors is monotone increasing.

Figure 7 shows the results of solving (29) numerically, with three different types of replication matrices. Several time steps are shown, with the asymptotic state clearly indicated as the outer envelope. As above the diagonal entries of $R$ are scaled to lie in [0, 1], we use $n = 6$, $\mu = .001$, and $b = .02$. The equation is solved using Strang's splitting method applied to the non-linear Kato-Trotter product formula. The parameter $E$ is selected so that $L + E$ has 11 positive eigenvalues. The initial data is $\boldsymbol{P}_0 = (1, \ldots, 1)$. Figure 7(a) shows the results with a random replication matrix. There appear to be only 10 distinct asymptotic quasispecies, though in fact the two left-most peaks have merged to form a plateau. The analysis presented in Section 4 is entirely born out in this numerical experiment. Figure 7(b) shows the results with a smooth monotone replication matrix. As predicted, the population has cascaded to form a single poorly localized quasispecies. In the final example, we use the single peak matrix, $R_{ii} = 1/(1 + d_H(S_i, S_0))$. As expected, this replication matrix produces one dominant quasispecies. Several smaller, but well localized quasispecies are also in evidence. The final set of figures shows the phase transition that occurs as the mutation rate is increased. For Figure 8, we solved (29) with $n = 7, b = .02$, and $\mu = .01, .1, 1, 10$. The transition from localized populations to delocalized populations is quite apparent. For each figure we use a different random replication matrix.

# A Mathematical Appendix

We consider models of the following general type:

$$\frac{d\boldsymbol{P}}{dt} = L\boldsymbol{P} - \boldsymbol{P} .* \boldsymbol{P} .* \boldsymbol{B} - \beta \langle \boldsymbol{P}, \mathbf{1} \rangle \boldsymbol{P} \tag{41}$$

where $\boldsymbol{B}$ is a pointwise positive super-harmonic vector and $\beta$ is a non-negative number. Here $L = R + M$, where $R$ is a diagonal matrix and $M$ is a matrix with zeroes on the diagonal and non-negative entries off the diagonal.

In order for such an equation to define a reasonable population model, it is necessary that it be positivity preserving. That is, if the initial data $\boldsymbol{P}(0)$ has non-negative entries, then $\boldsymbol{P}(t)$ is non-negative for all $t > 0$. The models of the type given in (41) have this property. This is established in two steps and uses the Kato-Trotter product formula and its non-linear generalization.

We first treat the linear part. The solution to the linear equation $\partial_t \boldsymbol{P} = L\boldsymbol{P}$ is given by

$$\boldsymbol{P}(t) = e^{tL}\boldsymbol{P}(0) = e^{t(R+M)}\boldsymbol{P}(0), \tag{42}$$

where, for a finite dimensional system, the matrix exponential is given by the usual formula, e.g.

$$e^{tL} = \sum_{j=0}^{\infty} \frac{(tL)^j}{j!}. \tag{43}$$

From this expression it is immediate that $e^{tM}$ has non-negative entries for every $t > 0$. Indeed for the models considered above $e^{tM}$ has positive entries for all $t > 0$. Because $R$ and $M$ do not commute, it does not follow immediately that $e^{t(R+M)}$ also has positive entries. To prove this we use the Kato-Trotter product formula, which states that

$$e^{t(R+M)} = \lim_{n\to\infty} \left[ e^{\frac{t}{n}R} e^{\frac{t}{n}M} \right]^n . \tag{44}$$

As the right hand side expresses $e^{t(R+M)}$ as a limit of products of matrices with non-negative entries, it follows that $e^{tL}$ also has non-negative entries. With a little more care we can show that, in fact, $e^{tL}$ has positive entries. Hence the linear model is positivity preserving. For a thorough discussion of positivity preserving operators see section XII.12 of [10].

In [12] a non-linear generalization of the Kato-Trotter formula is given for non-linearities including the type in (41). We first observe that the vector field defined on $\mathbb{R}^{2^n}$ by

$$X(\boldsymbol{P}) = -\boldsymbol{P}.*\boldsymbol{P}.*\boldsymbol{B} - \beta\langle \boldsymbol{P}, \mathbf{1}\rangle \boldsymbol{P} \tag{45}$$

is tangent to the coordinate hyperplanes $\{\boldsymbol{P} : P_j = 0\}$, and therefore the positive orthant, $\{\boldsymbol{P} : P_j > 0 \text{ for all } j\}$, is invariant under the flow generated by this vector field. From this and the fact that the right hand side in (45) is negative in the positive orthant, it follows that if we start with non-negative initial data, then the equation

$$\frac{d\boldsymbol{P}}{dt} = X(\boldsymbol{P}), \tag{46}$$

has a unique solution for all $t > 0$. Let $\mathscr{X}^t \boldsymbol{P}(0)$ denote the solution to (46) with initial data $\boldsymbol{P}(0)$. In [12] it is shown that the solution to (41) can be obtained as the following limit:

$$\boldsymbol{P}(t) = \lim_{n\to\infty} [e^{\frac{t}{n}L} \mathscr{X}^{\frac{t}{n}}]^n \boldsymbol{P}(0). \tag{47}$$

As $\mathscr{X}^t$ is positivity preserving and $e^{tL}$ is positivity improving, it follows that the equation in (41) is also positivity preserving. A small modification of this formula, useful for numerical simulations is called "Strang's splitting:"

$$\boldsymbol{P}(t) = \lim_{n\to\infty} [\mathscr{X}^{\frac{t}{2n}} e^{\frac{t}{n}L} \mathscr{X}^{\frac{t}{2n}}]^n \boldsymbol{P}(0). \tag{48}$$

We now consider the constraints imposed on the solution by the non-linearities. Assuming that $\boldsymbol{P}(t)$ is non-negative, it follows that there is a constant $M$ such that

$$\langle L\boldsymbol{P}, \mathbf{1}\rangle \leq M\langle \boldsymbol{P}, \mathbf{1}\rangle. \tag{49}$$

Thus, a non-negative solution to (41) satisfies the differential inequality:

$$\frac{d\langle \boldsymbol{P}, \mathbf{1}\rangle}{dt} \leq M\langle \boldsymbol{P}, \mathbf{1}\rangle - \beta\langle \boldsymbol{P}, \mathbf{1}\rangle^2. \tag{50}$$

This easily implies that if the initial population $\langle \boldsymbol{P}(0), \mathbf{1}\rangle < \beta^{-1}M$, then the total population never exceeds $\beta^{-1}M$. Moreover, if the population initially exceeds this value, then it decreases at time goes by. Combining this observation with the positivity preserving property, we deduce that, with non-negative initial data, the solution to (41) exists for all $t > 0$.

For the other non-linearity we use the hypothesis that $M$ has non-negative entries off the main diagonal. We suppose that $P_j(0) < B_j(0)$ for all $j$. Suppose that there were a $j_0$ and a first $t_0 > 0$, where $P_{j_0}(t_0) = B_{j_0}(t_0)$. In this case it would still be true that $P_j(t_0) \leq B_j(t_0)$, for all $j$. Hence, our assumption on $M$ and the fact that the remaining terms in $L$ are diagonal, would imply that

$$(L\boldsymbol{P}(t_0))_{j_0} \leq (L\boldsymbol{B})_{j_0}. \tag{51}$$

As the solution is non-negative this would imply the differential inequality

$$\left(\frac{dP_{j_0}(t_0)}{dt}\right) \leq (L\boldsymbol{B})_{j_0} - (\boldsymbol{B}. * \boldsymbol{B}. * \boldsymbol{B})_{j_0} < 0 \tag{52}$$

The last inequality is because $\boldsymbol{B}$ is assumed to be super-harmonic. But this contradicts the assumption that $P_{j_0}(t) < P_{j_0}(t_0)$, for $t < t_0$.

To sum up we have proved the following theorem:

**Theorem 1.** *If $\boldsymbol{P}(0)$ is non-negative, then the solution, $\boldsymbol{P}(t)$, to* (41) *exists for all time and remains non-negative. If $\langle \boldsymbol{P}(0), \mathbf{1}\rangle < \beta^{-1}M$, then this remains true for all time, and in any case remains bounded. If $\boldsymbol{B}$ is a positive super-harmonic vector, $L\boldsymbol{B} - \boldsymbol{B}. * \boldsymbol{B}. * \boldsymbol{B} < 0$, and $P_j(0) < B_j$, for all $j$, then this inequality remains true for all $t > 0$.*

It is likely that by treating the two non-linearities together, rather than separately as done above, more precise constraints could be obtained.

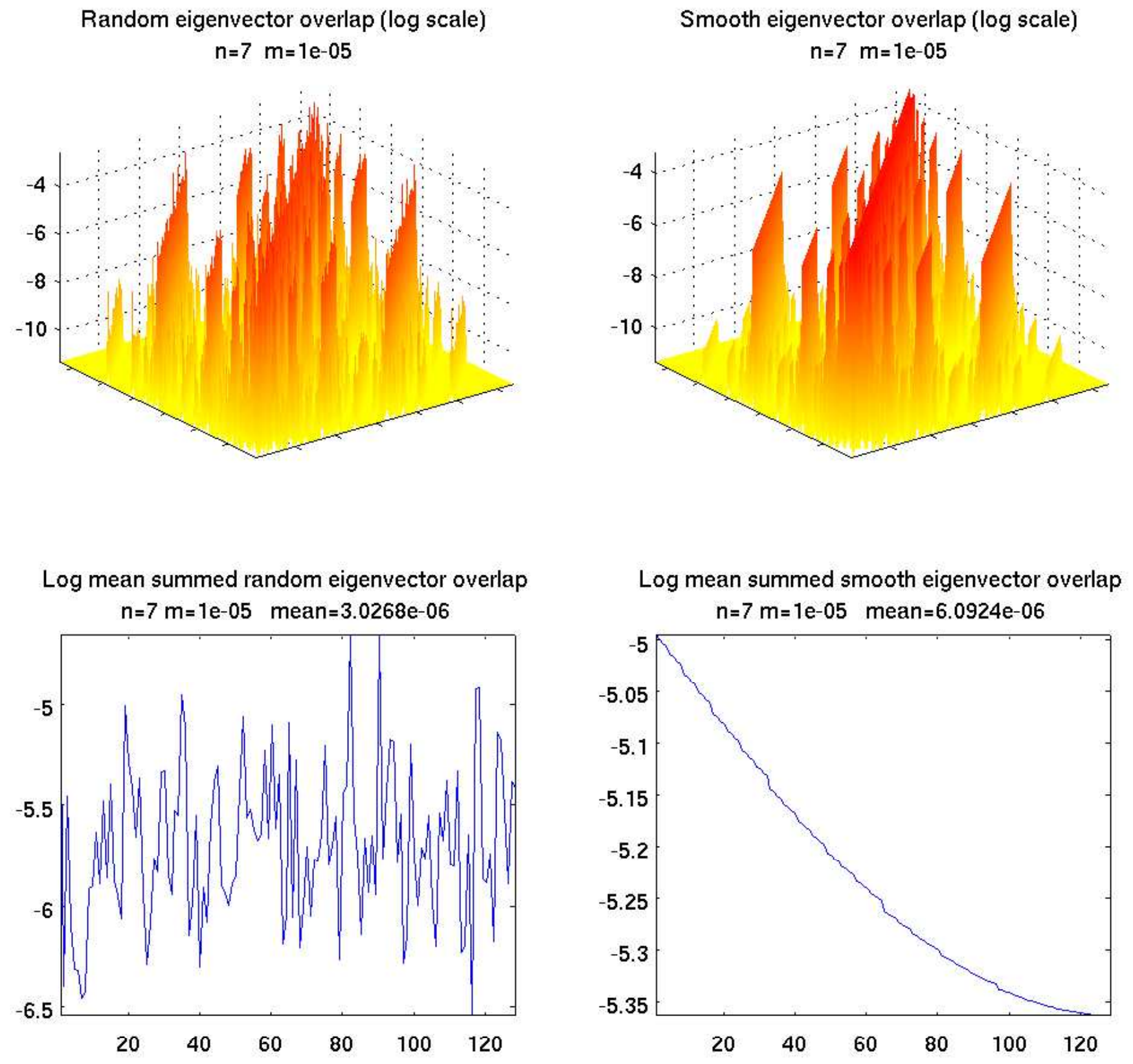

**Figure 1.** Trial with $n = 7$ and $\mu = 10^{-5}$. The smooth model has a monotonely increasing diagonal and the random model is uniformly distributed.

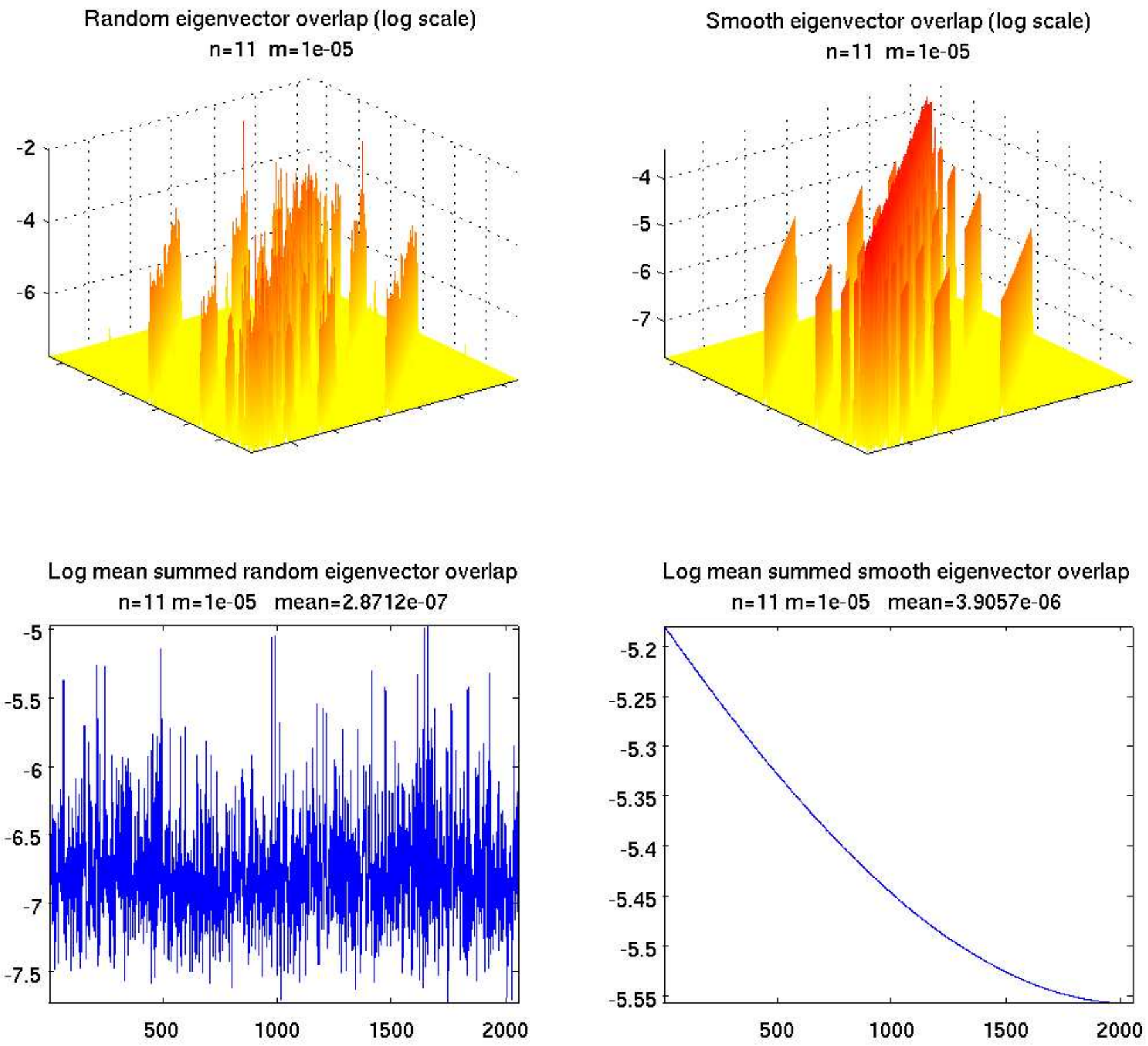

**Figure 2.** Trial with $n = 11$ and $\mu = 10^{-5}$. The smooth model has a monotonely increasing diagonal and the random model is uniformly distributed.

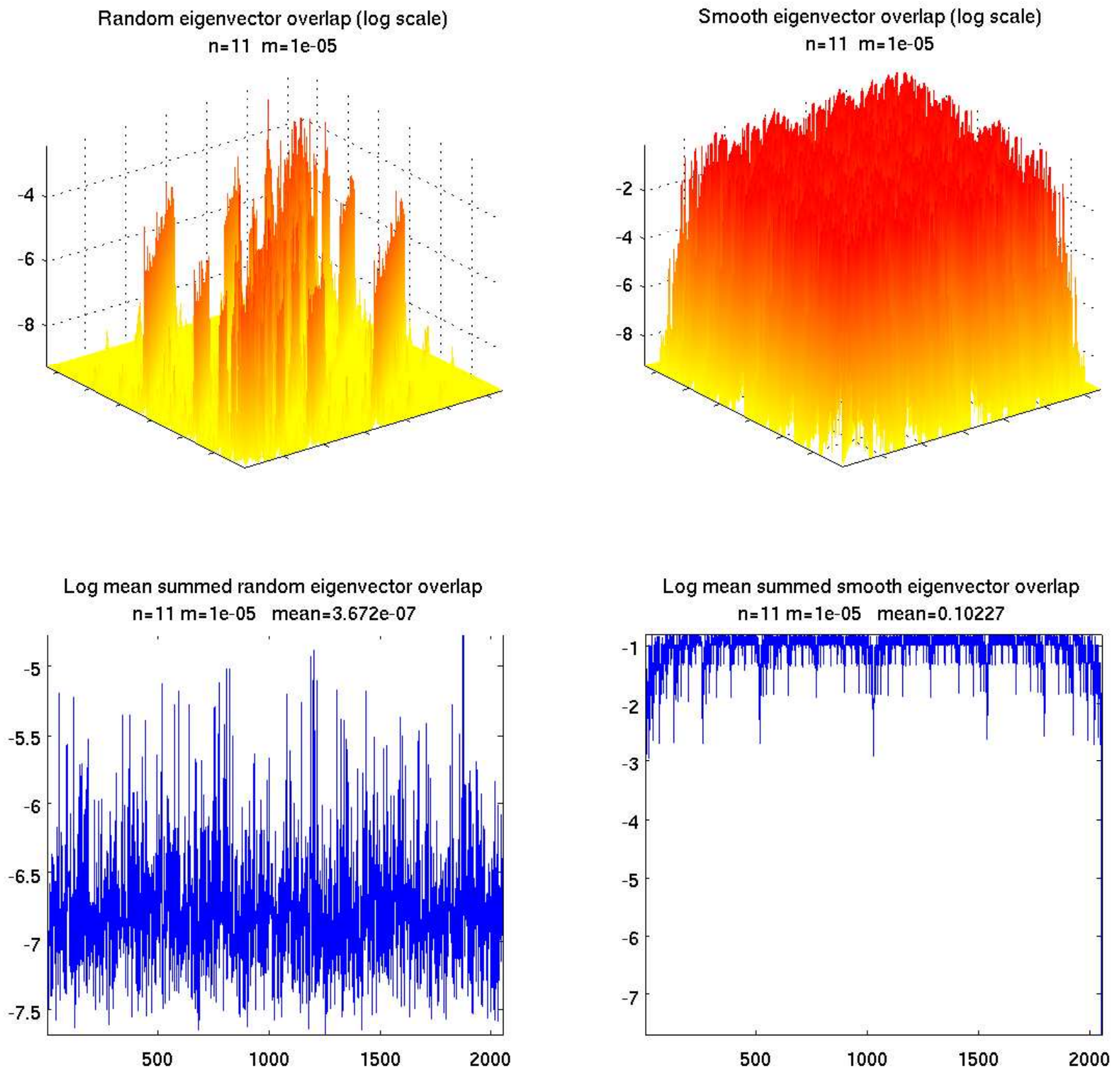

**Figure 3.** Trial with $n = 11$ and $\mu = 10^{-5}$. The smooth model has a single peak fitness landscape and the random model is uniformly distributed.

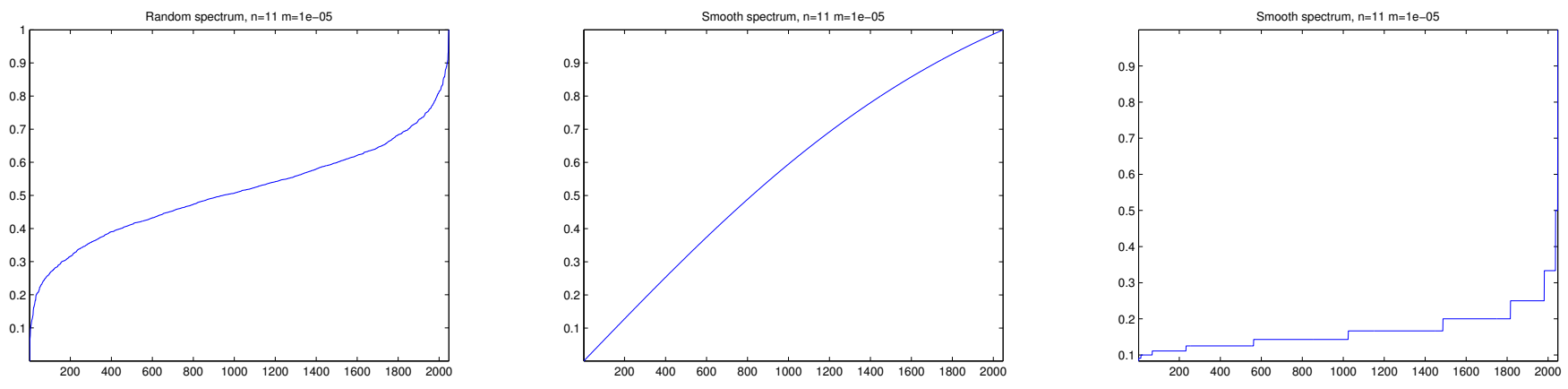

(a) Spectrum of a random model.

(b) Spectrum of a model with monotone replication matrix.

(c) Spectrum of a one-peak model.

**Figure 4.** The spectra of a random model, a monotone increasing model and a one-peak model with $n = 11$ and $\mu = 10^{-5}$.

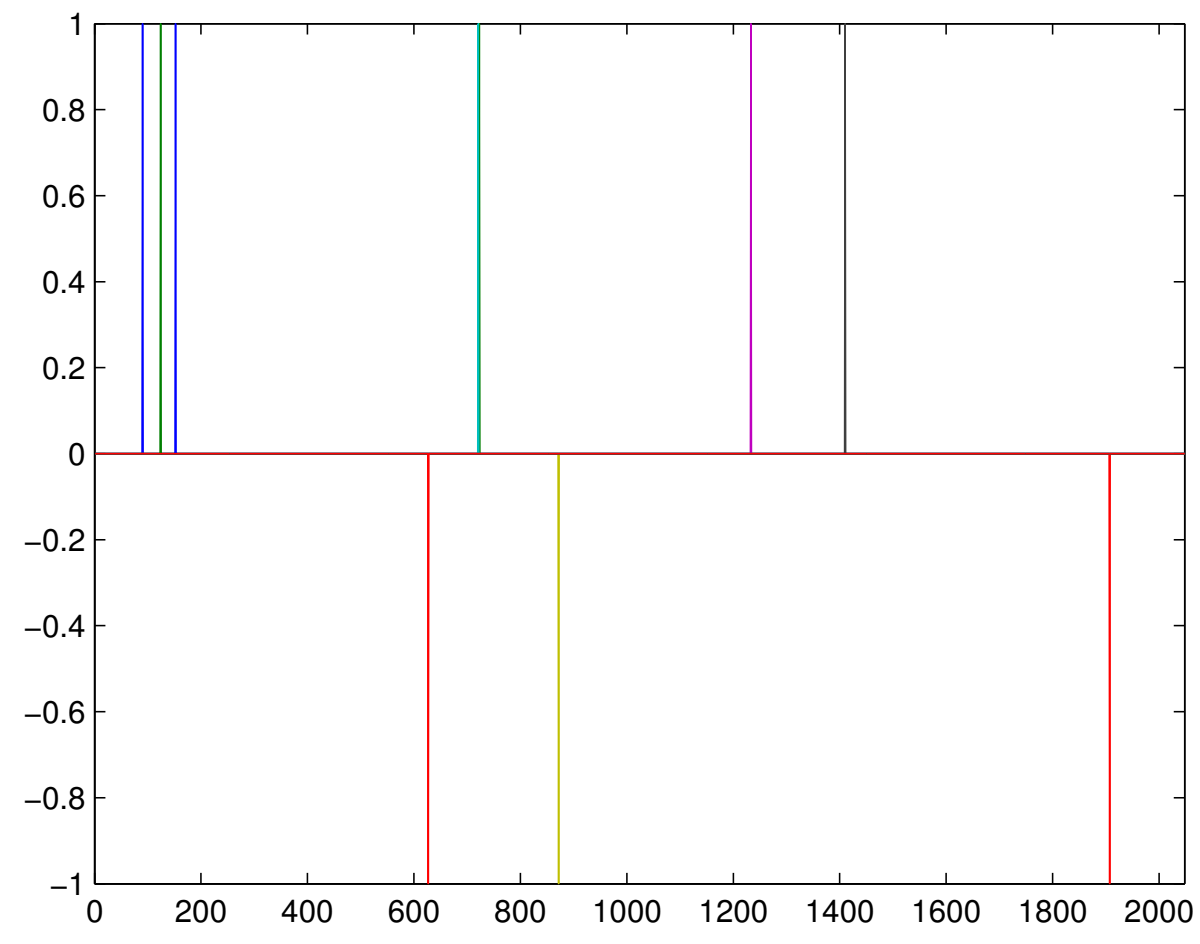

**Figure 5.** The first 10 eigenvectors (corresponding to the 10 largest eigenvalues) of a random model with $n = 11$ and $\mu = 10^{-5}$. There are, apparently, only 9 peaks as, eigenvectors 2 and 4 have considerable overlap.

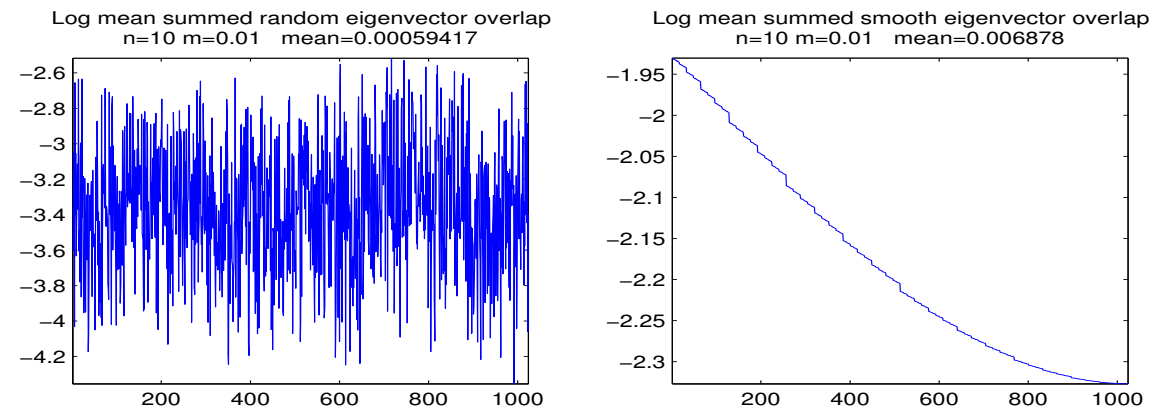

(a) $n = 10,\ \mu = 10^{-2}$.

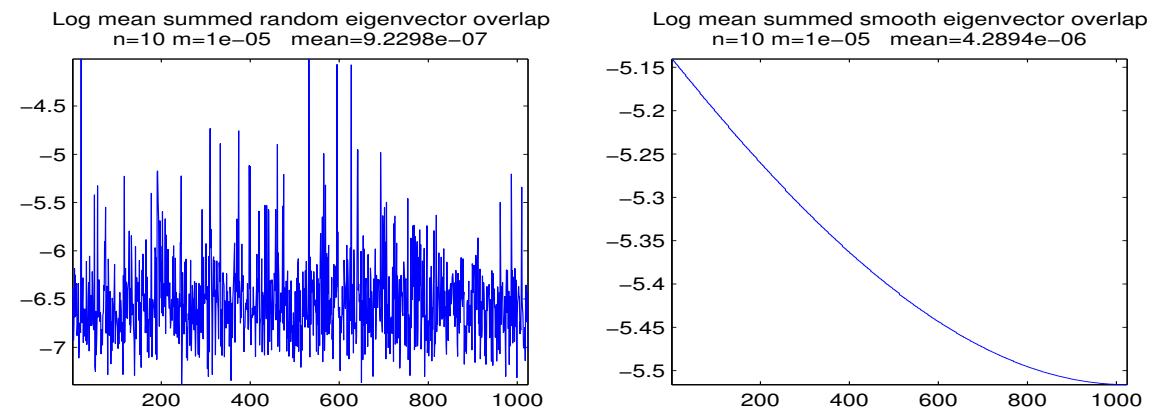

(b) $n = 10,\ \mu = 10^{-5}$.

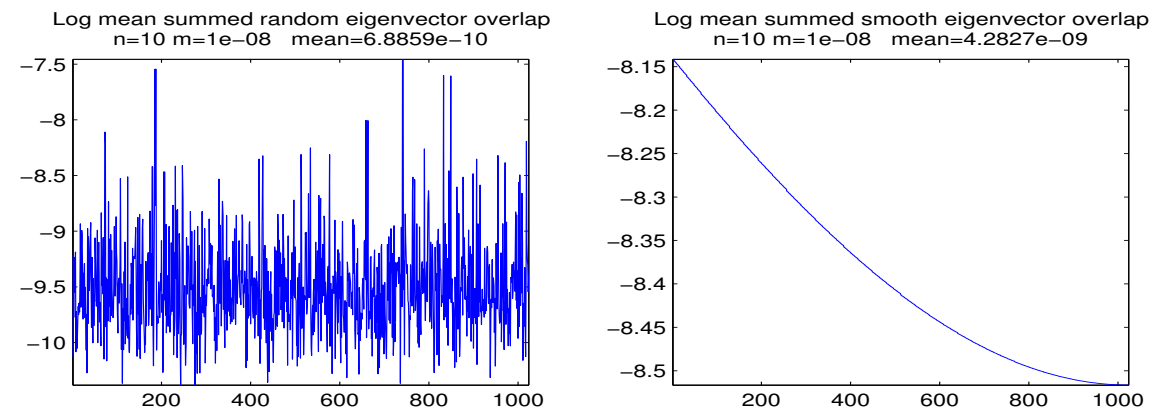

(c) $n = 10,\ \mu = 10^{-8}$.

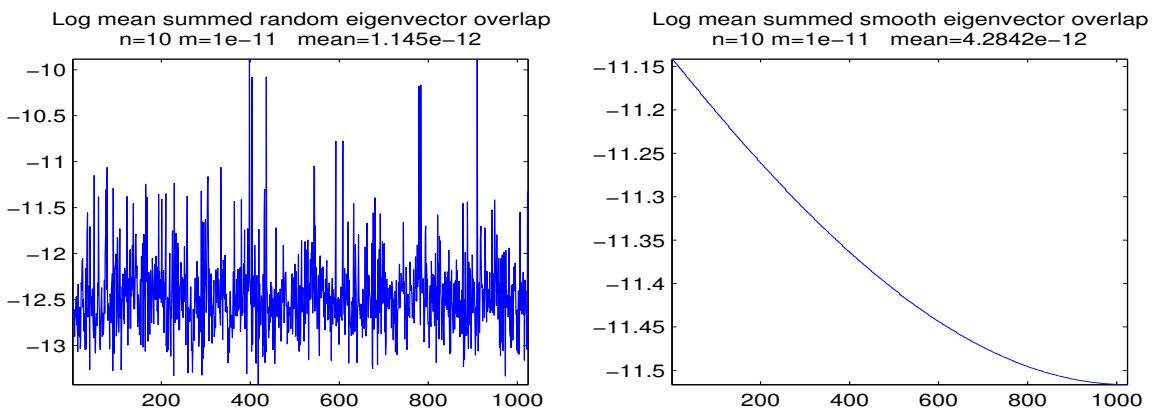

(d) $n = 10,\ \mu = 10^{-11}$.

**Figure 6.** The average summed overlap with different values of $\mu$, shown on a $\log_{10}$-scale. A random replication matrix is shown on the left and model with a monotone increasing diagonal on the right.

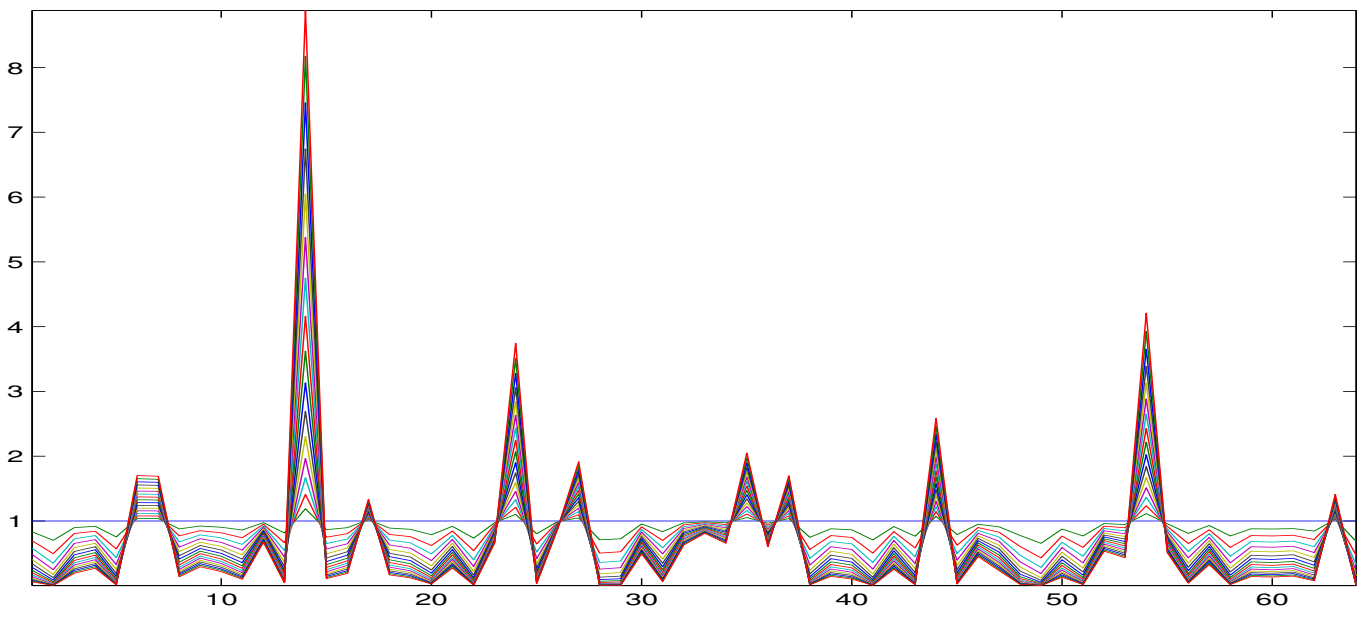

(a) Random replication matrix

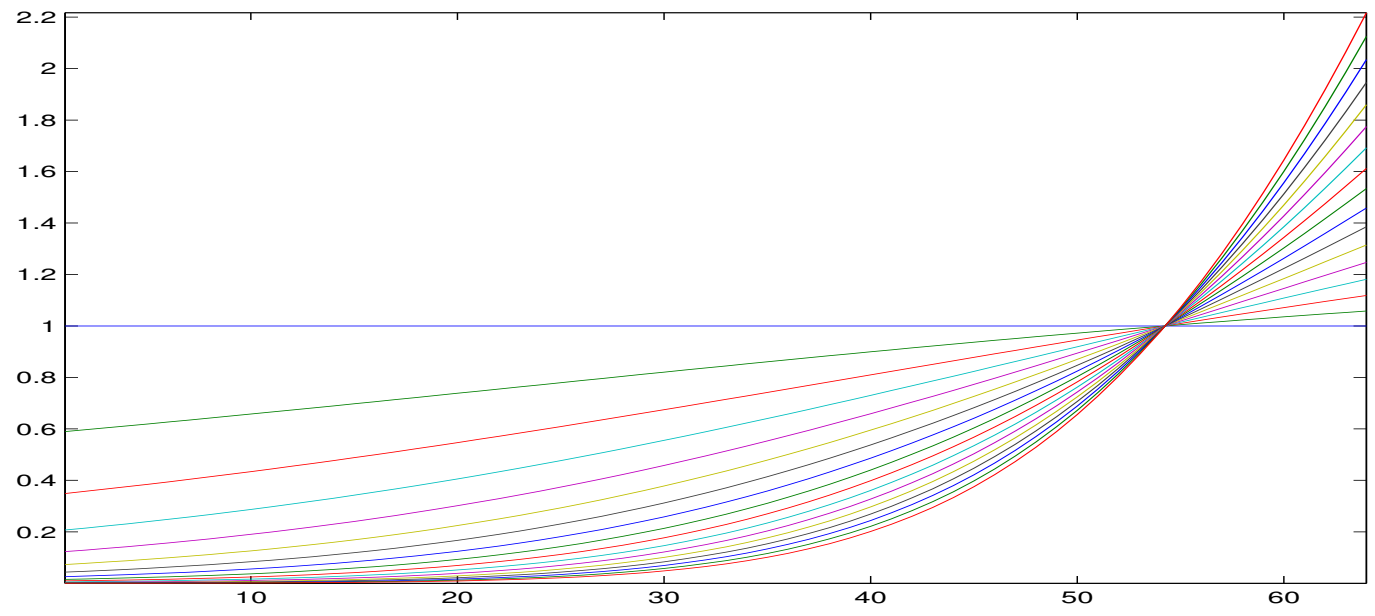

(b) Smooth monotone replication matrix

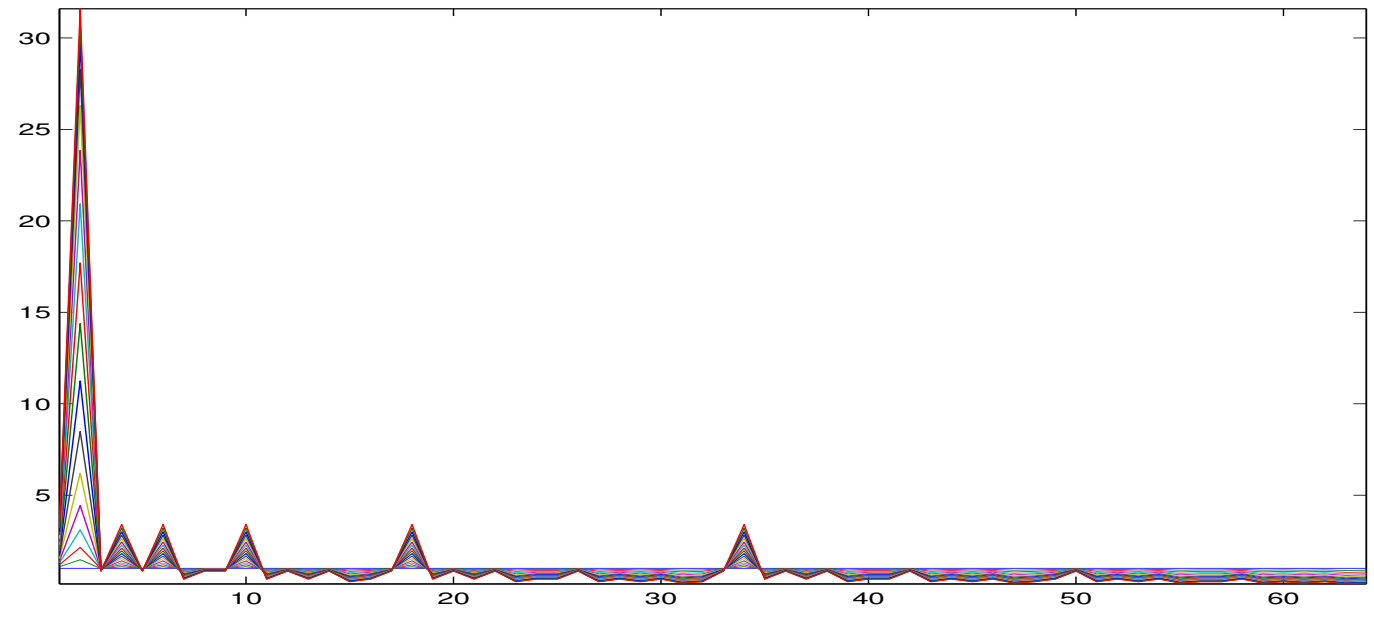

(c) Single peak replication matrix

**Figure 7.** The solution of the non-linear model, (29), with a variety of different replication matrices. The parameters are $n = 6, \mu = .001, b = .02$, and $\boldsymbol{P}_0 = (1, \ldots, 1)$. The energy $E$ is selected so that $L + E$ has 11 positive eigenvalues.

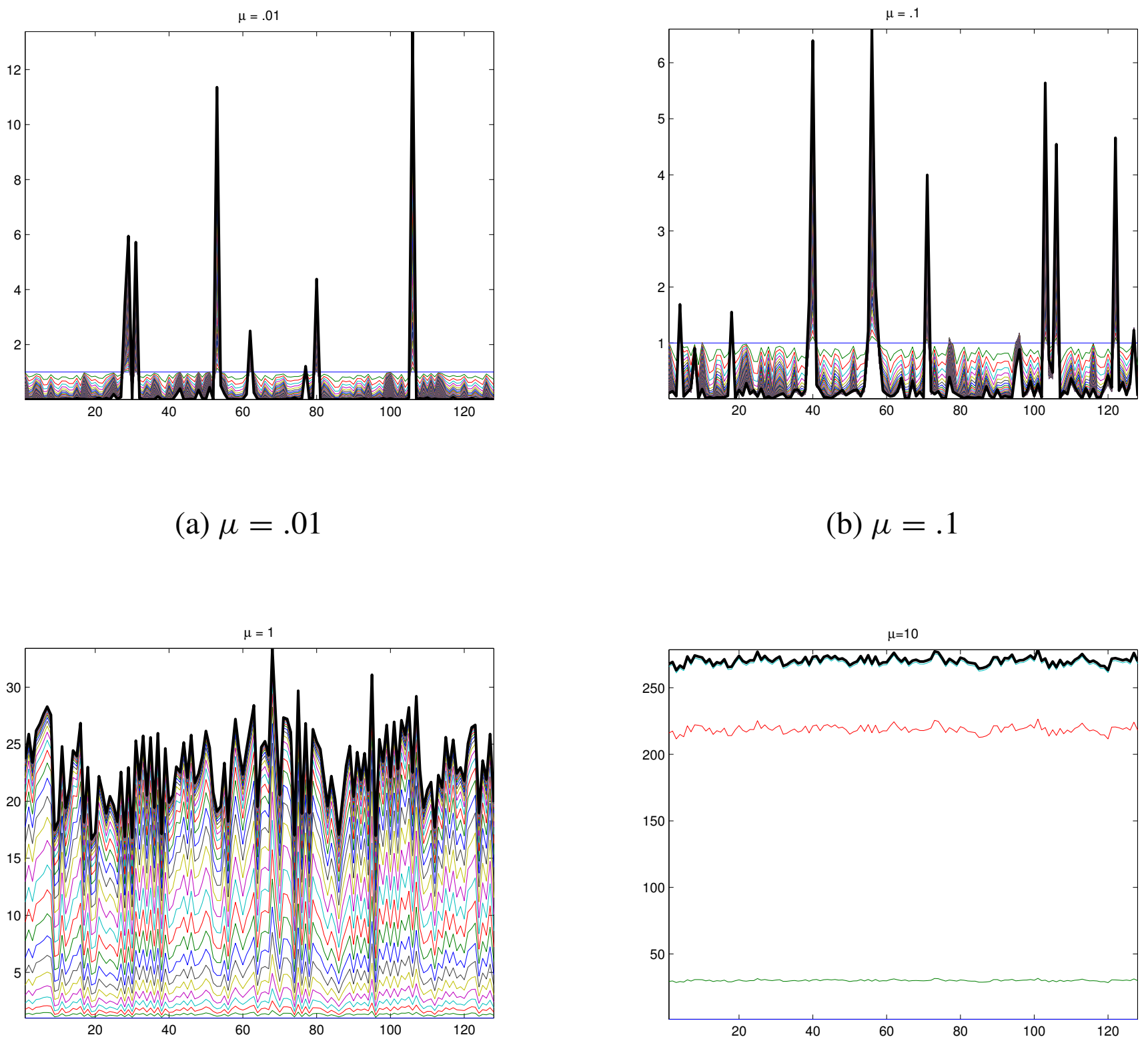

(a) $\mu = .01$ (b) $\mu = .1$

(c) $\mu = 1$ (d) $\mu = 10$

**Figure 8.** The solution of the non-linear model, (29), with a variety of different mutation rates. The parameters are $n = 7, b = .02$, $\boldsymbol{P}_0 = (1, \ldots, 1)$, and $\mu$ as indicated. The energy $E$ is selected so that $L + E$ has 10 positive eigenvalues.